\documentclass[groupaddress,twocolumn,prb,showpacs]{revtex4}
\usepackage[dvips]{graphicx}
\usepackage{amsmath}

\def\zro2{ZrO$_2$}
\def\hfo2{HfO$_2$}
\def\sio2{SiO$_2$}

\begin{document}

\title{Structural and dielectric properties of amorphous \zro2 and \hfo2}
\author{Davide Ceresoli}
\altaffiliation[Present address: ]{Scuola Internazionale Superiore di Studi
Avanzati (SISSA) and DEMOCRITOS, via Beirut 2-4, I-34014 Trieste, Italy}
\affiliation{Department of Physics and Astronomy, Rutgers University,
136 Frelinghuysen Road, Piscataway, New Jersey 08854, USA}
\author{David Vanderbilt}
\affiliation{Department of Physics and Astronomy, Rutgers University,
136 Frelinghuysen Road, Piscataway, New Jersey 08854, USA}
\date{June 13, 2006}

\begin{abstract}

Zirconia (\zro2) and hafnia (\hfo2) are leading candidates for replacing
\sio2\ as the gate insulator in CMOS technology. Amorphous versions of these
materials ($a$-\zro2\ and $a$-\hfo2) can be grown as metastable phases on
top of a silicon buffer; while they tend to recrystallize during subsequent
annealing steps, they would otherwise be of considerable interest because
of the promise they hold for improved uniformity and electrical passivity.
In this work, we report our theoretical studies of $a$-\zro2 and $a$-\hfo2 by
first-principles density-functional methods. We construct realistic
amorphous models using the ``activation-relaxation'' technique (ART) of Barkema
and Mousseau.  The structural, vibrational, and dielectric properties of
the resulting models are analyzed in detail. The overall average dielectric
constant is computed and found to be comparable to that of the monoclinic
phase.
\end{abstract}

\maketitle


\section{Introduction}\label{sec:intro}

In recent years, a major thrust of applied development in the semiconductor
industry has been the search for materials that could replace \sio2\ as
the gate dielectric in CMOS technology.  Conventional scaling of
\sio2\ would require a gate-dielectric thickness shrinking below 1 nm
in the near future, according to the International Technology Road Map
for Semiconductors.\cite{itrs}  In this
regime, conventional thermally-grown \sio2\ is expected to fail
because of issues of tunneling leakage current and reliability.
There are therefore strong incentives to identify
materials with dielectric constant $\epsilon$ (or $K$)
much larger than that of \sio2\ (at $\epsilon$=3.9), as these could
be grown as thicker films while still providing the
needed capacitance.

Among the most promising materials are \zro2\ and \hfo2,
which do have much higher dielectric constants and some other
positive features as well (e.g., chemical stability).
However, one of the great advantages of \sio2\ has
been the fact it forms an amorphous oxide ($a$-\sio2), thus
allowing it to conform to the substrate with enough freedom to
eliminate most electrical defects at the interface.
On the other hand, \zro2\ and \hfo2\ (which are very similar in many
of their physical and chemical properties) are refractory materials,
their melting temperature being 2988\,K and 3085\,K respectively.
They are not good glass-formers; $a$-\zro2, for example, has been shown
to recrystallize during growth if the growth temperature is too
high (with tetragonal and monoclinic phases starting to appear above
$\sim$\,500\,$^\circ$C and $\sim$\,700\,$^\circ$C respectively).\cite{zhu04}
Thus, while \zro2\ and
\hfo2 can be grown as metastable amorphous phases on Si using
low-temperature deposition techniques, films of this type
unfortunately tend to recrystallize during the subsequent annealing
steps that are required in current industrial fabrication
processes.

Nevertheless, it is possible that admixing (alloying)
with Si, Al, N, or other chemical constituents, or other strategies
yet to be identified, may help to mitigate the recrystallization
problem and stabilize the amorphous phase.\cite{zhu04}
In any case, it may be advantageous to study amorphous structures
as a first step in understanding why the recrystallization is
facilitated.
With these motivations, we have embarked on a theoretical study of the
structural and dielectric properties of
amorphous zirconia and hafnia ($a$-\zro2 and $a$-\hfo2).
In earlier work,\cite{ZCV04} we generated models of
$a$-ZrO$_2$ using an {\it ab-initio} molecular dynamics approach
in a plane-wave pseudopotential framework, and studied
their structural and dielectric properties.  In a subsequent
publication,\cite{vanderbilt05} we described an alternate and
more efficient method of generating amorphous structures based
on the activation-relaxation technique (ART) of Barkema and
Mousseau,\cite{ART} and presented a few preliminary results on
amorphous ZrO$_2$ obtained using this approach.  These calculations
were carried out using a local-orbital basis approach embodied
in the SIESTA code package.  \cite{SIESTA}  In the present work,
we systematically use SIESTA for all of the reported
calculations.  We first briefly review our implementation of the
ART approach, and then embark on a systematic description of the
generated $a$-\zro2 and $a$-\hfo2 structures and a comparison of
their computed electronic, structural, and dielectric properties.

The paper is organized as follows.  In Secs.~II and III we describe the
details of our electronic-structure calculations and of our procedure
for generating amorphous models, respectively.  We then describe the
results of our calculations for ZrO$_2$ in Sec.~IV and for HfO$_2$
in Sec.~V.  We also present some results concerning the Born
dynamical effective charges and the dielectric activity
activity for both systems in Sec.~VI.  We then finish with a conclusion
in Sec.~VII.

\section{Computational details}\label{sec:computational}

We performed {\it ab-initio} density-functional theory (DFT) calculations
\cite{DFT} in the local-density approximation (LDA) \cite{LDA}
using a local-orbital expansion of the
Bloch wavefunction as implemented in the SIESTA code.\cite{SIESTA}
The use of localized-orbital basis allows the retention of good accuracy in
the DFT calculations at a reduced computational cost relative to plane-wave
codes.  A cutoff of 150 Ry was used for the expansion of the charge density,
and all calculations were performed using $\Gamma$-point Brillouin zone
sampling only.
The dynamical matrix and Born effective charges were
computed using the SIESTA via a finite-difference approach in which each
atom was displaced by $\pm$\,0.05\,\AA\ in the three cartesian directions
and the forces and polarization changes were computed.

Two kind of basis sets were employed in the calculations. A minimal
single-$\zeta$ basis set was used for the amorphization procedure, which
required a large number of structural relaxations, while a larger
triple-$\zeta$ basis set was used for the structural relaxation of the
final structure and for the calculation of the lattice dielectric
constant and of the infrared activity.  Such a two-stage procedure
(amorphization with a lower-level theory followed by annealing
and relaxation with a higher-level theory) is a natural way of
establishing a trade off between computational cost and accuracy;
it has many precedents in the literature, as for example the
generation of an amorphous structure by the use
of empirical potentials followed by a first-principle annealing
run in the work of Sarnthein {\it et al.}\cite{sarnthein95}

Because we were concerned that
the minimal basis might not accurately reproduce the delicate interplay
between ionic and covalent bonding in \zro2\ and \hfo2, the
parameters for the minimal basis set were optimized following the
prescription of Ref.~\onlinecite{javier} in order to ensure the
correct energy ordering of the cubic, orthorhombic and monoclinic
phases of \zro2\ and \hfo2.
We then obtain $E_\mathrm{cub}-E_\mathrm{tet}=$~62~meV and
$E_\mathrm{tet}-E_\mathrm{mono}=$~89~meV for \zro2,
to be compared with values of
44~meV and 45~meV, respectively, computed using plane-wave methods and
reported in Ref.~\onlinecite{zhao02}.
Here $E_\mathrm{cub}$, $E_\mathrm{tet}$, and $E_\mathrm{mono}$ are the
energies per formula unit of the cubic, tetragonal, and monoclinic phases,
respectively of \zro2.  Similar results are found for \hfo2.
The relaxed structural parameters are also found to agree well.
Thus, despite their simplicity, our optimized minimal basis sets for
\zro2\ and \hfo2 are able to capture the essentials of the structural
energetics of the three crystalline phases.

\section{Amorphization procedure}\label{sec:amorphization}

In a previous study,\cite{ZCV04} we carried out {\it ab-initio}
molecular dynamics (MD) simulations of a 96-atom supercell in a
melt-and-quench fashion in order to generate a structural model for
$a$-\zro2. Because of the short time interval
accessible to simulations, the cooling rate (3.4$\times$10$^{14}~$K/s)
was far beyond the fastest cooling rate that can be
obtained experimentally by pulsed laser techniques.
The fast cooling rate does not allow long-time-scale relaxation of system,
which might be important in the case of relatively poor glass formers such
as \zro2\ and \hfo2.

In order to see if the fast cooling rate could bias the resulting amorphous
structure, we generated an independent amorphous sample\cite{vanderbilt05}
by using the ART event-based structural-evolution approach.\cite{ART}
An event-based method leads to an accelerated dynamics and to a better
sampling of long-time-scale modes in glasses.
Rather than following the irrelevant details of the atomic motions
as atoms vibrate back and forth about their average positions for long
periods between activated events, ART
focuses on simulating jumps over the barriers that separate
the different basins of attraction of different local minima in the energy
landscape.  Thus, it requires a given computational effort per activated hop
instead of per vibrational period. For disordered systems, which tend to show
slow evolution, these two time scales may differ by many orders of magnitude,
and this event-based technique therefore allows a much faster simulation.

We implemented the ART method as a driver for the {\it ab-initio} code
SIESTA.\cite{SIESTA} The flowchart of our implementation of the ART
method can be found in Ref.~\onlinecite{vanderbilt05}.  As reported there,
we found that, starting from a 96-atom supercell made
from the perfect cubic fluorite crystal structure, a number of
Monte-Carlo (MC) trials equal to 5 times the number of atoms (i.e.,
$\sim\,$500) was sufficient to produce a good amorphous structure.
In our case, a MC temperature of 3000\,K produced an acceptance ratio of
13\%. Moreover, we found that during the first 50 MC trials, the acceptance
ratio was higher ($\sim\,$40\%) and dropped down in the subsequent trials.
We did not attempt to perform longer simulations at a lower MC temperature
in order to ``anneal'' the system further.

\section{Results for $\mathbf{ZrO}_2$}\label{sec:zro2}

An important issue is the density of the amorphous phase, which, to our
knowledge, is not accurately known experimentally. This issue was explored
in our previous papers\cite{ZCV04,vanderbilt05} and we found
that among the initial guesses, a density between 4.86~g/cm$^3$ and
5.32~g/cm$^3$ generates a robust amorphous structure for
\zro2.\cite{explan-seven}

\begin{figure}
  \includegraphics[width=0.65\columnwidth]{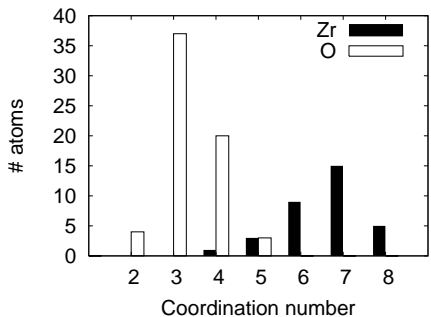}
  \caption{$a$-\zro2\ distribution of coordination numbers resulting from
  the activation-relaxation (ART) simulation. Zr and O atoms are indicated by
  filled and open bars, respectively.}
  \label{fig:zirconia_coord}
\end{figure}

The ART simulation was performed at a constant volume corresponding to a
density of 5.32~g/cm$^3$. A snapshot of the system is essentially
indistinguishable, at the visual level, from that of the melt-and-quench
MD-generated system of Ref.~\onlinecite{ZCV04}.
The corresponding distribution of coordination numbers is shown in
Fig.~\ref{fig:zirconia_coord}. We find a prevalence of 7- and
6-coordinated Zr over 8-coordinated Zr and a prevalence of 3- and
4-coordinated oxygens, similar to what was found for the MD-generated
structure.  (For comparison, recall that the monoclinic structure
has 7-fold cations and equal numbers of 3- and 4-fold anions.)

\begin{figure}[b]
  \includegraphics[width=0.65\columnwidth]{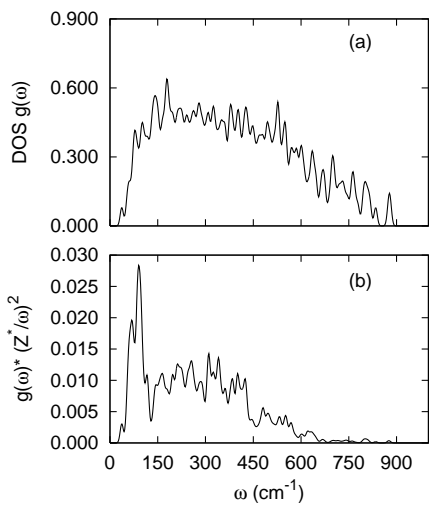}
  \caption{(a) $a$-\zro2\ phonon density of states (DOS) vs frequency.
  (b) $a$-\zro2\ infrared activity (phonon DOS weighted by
  $\widetilde{Z}^2_\lambda/\omega^2_\lambda$) vs frequency.}
  \label{fig:zirconia_spectrum}
\end{figure}

One may wonder if such a spread of coordination numbers is actually due to
intrinsic coordination defects such as dangling bonds or non-bridging oxygens
(which are often present in other oxides such as \sio2). We monitored
the electronic density of states at every MC accepted move and we
found that occasionally a defective structure (i.e., with some levels in the
gap) is accepted. We also found, however, that the
coordination defects are promptly
saturated in subsequent MC moves. The final amorphous sample is a good
insulator with an electronic gap of $\sim\,$3.4\,eV.

Furthermore, the calculated phonon spectrum, shown in
Fig.~\ref{fig:zirconia_spectrum}, is found to
extend over approximately the same range of frequencies (50-800\,cm$^{-1}$)
and to show features similar to those of the sample obtained by the
melt-and-quench
MD simulation. The computed Born effective charges are slightly larger on
average, Z$^*$(Zr)=+5.08 and Z$^*$(O)=$-$2.54.
Our lattice dielectric tensor is
\begin{equation}
  \epsilon_\mathrm{latt} = \left(
  \begin{array}{rrr}
   \phantom{-}17.9 & -0.7 & \phantom{-}0.2 \\
   -0.7 & \phantom{-}17.5 & -0.5 \\
   \phantom{-}0.2 & -0.5 & \phantom{-}14.1
  \end{array}\right)
\end{equation}
yielding an average dielectric constant of 16.5, compared to the value of 17.6
of the MD-generated model, due to the fact that the vibrational spectrum is
shifted to slightly higher frequencies.
Assuming a value of $\epsilon_\infty$\,=\,4.6 for the
high-frequency dielectric constant,\cite{ZCV04} the
static dielectric constant yields a value of $\sim$\,21, in good agreement
with experimental values~\cite{devine01} and with the previous
calculation.\cite{ZCV04}

\section{Results for $\mathbf{HfO}_2$}\label{sec:hfo2}

\begin{figure}
  \includegraphics[width=0.65\columnwidth]{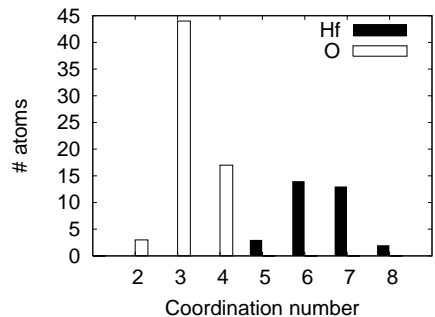}
  \caption{$a$-\hfo2\ distribution of coordination numbers resulting from
  the activation-relaxation (ART) simulation. Hf and O atoms are indicated by
  filled and open bars, respectively.}
  \label{fig:hafnia_coord}
\end{figure}

\begin{figure}[b]
  \includegraphics[width=0.65\columnwidth]{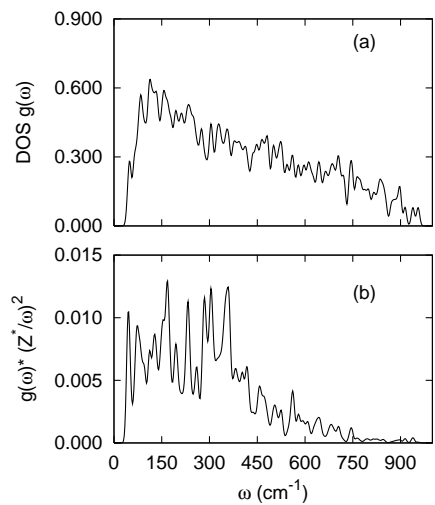}
  \caption{(a) $a$-\hfo2\ phonon density of states (DOS) vs frequency.
  (b) $a$-\hfo2\ infrared activity (phonon DOS weighted by
  $\widetilde{Z}^2_\lambda/\omega^2_\lambda$) vs frequency.}
  \label{fig:hafnia_spectrum}
\end{figure}

The amorphization of \hfo2\ was carried out in a manner very
similar to that used for its twin oxide \zro2. For \hfo2, however,
we avoided the tedious procedure of performing
independent melt-and-quench simulations in order to find a suitable
density that can sustain an amorphous structure. Instead, we started the
ART simulation at the same volume as for \zro2, 38.45\,\AA$^3$ per
unit formula (corresponding to an \hfo2\ mass density of 9.09\,g/cm$^3$).
Then, every 10 accepted MC moves, we performed a relaxation of the
supercell volume. The final amorphous \hfo2\ sample had a density of
9.39\,g/cm$^3$ (corresponding to 37.22\,\AA$^3$ per unit formula,
3.2\,\% smaller than that of \zro2). Experimentally, the unit cell volumes of
the monoclinic and cubic phases are $\sim$2\% larger in \hfo2\ than in
the corresponding \zro2\ phases.\cite{zhao02}

The distribution of coordination numbers of the resulting $a$-\hfo2\ sample
is shown in Fig.~\ref{fig:hafnia_coord}. The average coordination number
shows a prevalence of 7- and 6-coordinated Hf over 8-coordinated Hf, and a
prevalence of 3- and 4-coordinated oxygens, as in the monoclinic structure.
Compared to $a$-\zro2\ (Fig.~\ref{fig:zirconia_coord})
there is a slightly higher prevalence of 3-coordinated oxygens, while
the distribution of 6- and 7-coordinates cations is broadly similar.
However, it is doubtful whether the small differences between
\zro2\ and \hfo2\ that are visible in Figs.~\ref{fig:zirconia_coord}
and \ref{fig:hafnia_coord} are statistically significant.

The generated $a$-\hfo2\ sample is found to be insulating, with a gap
of $\sim$\,3.8\,eV.  Its calculated phonon spectrum, shown in
Fig.~\ref{fig:hafnia_spectrum}, extends over approximately the
same range of frequencies (50-800\,cm$^{-1}$) as for $a$-\zro2.
Compared to Fig.~\ref{fig:zirconia_spectrum}(b), the infrared activity
for \hfo2\ in Fig.~\ref{fig:hafnia_spectrum}(b) shows a broader feature
in the 50-400\,cm$^{-1}$ frequency range.  The computed Born effective
charges are found to be smaller, on average, compared to those
of $a$-\zro2: Z$^*$(Hf)=+4.8 and Z$^*$(O)=$-$2.4.
Our computed lattice dielectric tensor
\begin{equation}
  \epsilon_\mathrm{latt} = \left(
  \begin{array}{rrr}
   \phantom{-}23.4 & -5.1 & -1.2 \\
   -5.1 & \phantom{-}14.4 & \phantom{-}0.2 \\
   -1.2 & \phantom{-}0.2 & \phantom{-}12.6
  \end{array}\right)
\end{equation}
yields an average lattice dielectric constant of 16.8.
Assuming a value of $\sim$\,5
for the high-frequency dielectric constant ($\epsilon_\infty$),\cite{zhao02}
the static dielectric constant yields a value of $\sim$\,22, confirming again
the striking similarity between $a$-\zro2\ and $a$-\hfo2.

\begin{figure}
  \includegraphics[width=0.65\columnwidth]{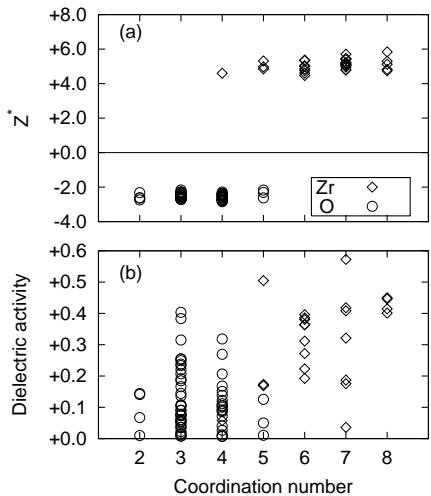}
  \caption{(a) Scatter plot of isotropically averaged atomic $Z^*$
  values (vertical axis) vs atom type and coordination number (horizontal
  axis) for $a$-\zro2.
  Circles and diamonds denote O and Zr atoms, respectively. (b)
  Same but with ``dielectric activity''
  (see Eq.~(5) of Ref.~\onlinecite{ZCV04}) plotted vertically.}
  \label{fig:zirconia_zstar_activ}
\end{figure}

\section{Born charges and dielectric activity}

Clearly it is desirable to understand more fully the various contributions to
the lattice dielectric response of the amorphous forms of \zro2\ and \hfo2.
To this end, we decomposed various lattice properties by ``atom type''
(that is, by chemical species and coordination number) in the hope that such
an analysis may provide further insight into our numerical results.
This type of analysis was explained in detail and applied to
the $a$-\zro2\ sample generated using the MD melt-and-quench approach
in Ref.~\onlinecite{ZCV04}.

\begin{figure}
  \includegraphics[width=0.65\columnwidth]{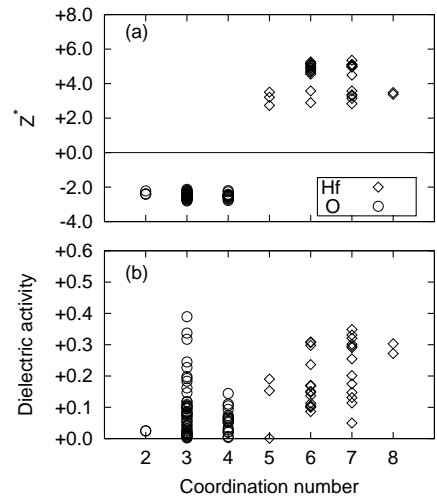}
  \caption{(a) Scatter plot of isotropically averaged atomic $Z^*$
  values (vertical axis) vs atom type and coordination number (horizontal
  axis) for $a$-\hfo2.
  Circles and diamonds denote O and Hf atoms, respectively. (b)
  Same but with ``dielectric activity''
  (see Eq.~(5) of Ref.~\onlinecite{ZCV04}) plotted vertically.}
  \label{fig:hafnia_zstar_activ}
\end{figure}

Our results for the ART-generated samples are presented in
Figs.~\ref{fig:zirconia_zstar_activ} and
Fig.~\ref{fig:hafnia_zstar_activ}.  In the top panel of each figure, we
report the Born effective charges as a function of the chemical
species and coordination number. In both cases, the Born effective
charges tend to be surprisingly independent of coordination
number, and quite similar to the values found in the crystalline
phases.\cite{zhao02}  However, a mild tendency of the effective charge to
increase with increasing coordination number of the cation is
visible especially for Zr; it is also present for Hf but is
partially obscured by larger fluctuations of the effective charges
in that case.

The atom-resolved dielectric activity, defined in Eq.~(5) of
Ref.~\onlinecite{ZCV04}, is presented in
Figs.~\ref{fig:zirconia_zstar_activ}(b) and \ref{fig:hafnia_zstar_activ}(b).
This is essentially a measure of the contribution of
each type of atom to the lattice dielectric constant.  Some trends
are visible, such as a stronger contribution by 3-fold than by
4-fold oxygens, and a tendency to have larger contributions from
higher-coordinated cations.  However, even within any one atom
type, there is a surprising degree of variation, with some
atoms contributing strongly and others contributing very little.
Since the effective charges do not have nearly such a large
variation, we infer that the differences must arise because of
differences in the force-constant matrix elements.  That is, some
atoms may be regarded as being strongly anchored in place so that they
contribute only weakly to the dielectric activity, while
others participate in one or more soft modes and contribute
strongly.

\section{Conclusions}\label{sec:conclusions}

In conclusion, we have applied {\it ab-initio} electronic structure
methods to study the lattice dynamics and dielectric properties of
amorphous high-$K$ materials $a$-\zro2\ and $a$-\hfo2.  We
used the ART event-based technique to generate structural models of
these amorphous materials.  These 96-atom supercell models display
a distribution of coordination numbers (mostly 3- and 4-fold for
oxygen and 6-, 7- and 8-fold for Zr and Hf) but nonetheless remain
insulating with a robust gap.  The full force-constant matrix was
computed for each supercell model, and the phonon density of states
was calculated.  In addition, the full Born-charge tensors were also
computed for each model, and when combined with the force-constant
information, the lattice dielectric response was obtained.  The
total dielectric constant is computed to be $\sim$\,22 for both
materials, comparable to that of the monoclinic phase.
Finally, the Born charges and the contributions to the dielectric
activity were further analyzed by atom type.  The effective charges
are relatively uniform and are roughly similar to those of the
crystalline phases, whereas there are strong variations in the 
individual atomic contributions to the dielectric activity resulting
from variations in soft-mode participation.

Several caveats are in order.  First, while our ART procedure can be
regarded as corresponding to a slower quench than in the previous
MD-based calculations,\cite{ZCV04} it is still very fast relative to
any experimental quenching time scale, so that our models might have
more defects than the experimental systems of interest.
Second, real amorphous systems are generally grown
at sufficiently low temperature to avoid crystallization, and not
quenched from a melt.  Thus, there could be significant differences
in the resulting structures (for example, a greater tendency to
void formation in the experimental samples).  Third, our samples
are perfectly stoichiometric and impurity-free, which will
not generally be true of experimental samples.  Finally, while a
statistical analysis of an ensemble of ART-generated samples would
be very desirable, the computational burden needed to carry out
the above analysis for even a single sample is quite demanding.
Nevertheless, we believe that the present results reveal the broad
qualitative features to be expected for these amorphous phases,
and constitute an important step forward on the road to a better
understanding of these important materials.

\bigskip

\acknowledgments

D.C. acknowledges J. Junquera for useful discussions and advice on basis
set optimization. This work was supported by NSF Grant DMR-0233925.


\end{document}